\documentclass[english,12pt,a4paper]{article}

\usepackage[T1]{fontenc}
\usepackage[latin9]{inputenc}

\usepackage[a4paper]{geometry}
\usepackage{setspace}
\usepackage{amsmath}
\usepackage{graphicx}
\usepackage[unicode=true]{hyperref}
\usepackage{cite}

\renewcommand{\eqref}[1]{(\ref{#1})}
\newcommand{\figref}[1]{Fig.~\ref{#1}}

\newcommand{\rmd}{\mathrm{d}}

\title{Bessel--like birth--death process}
\author{Vygintas Gontis, Aleksejus Kononovicius}
\date{Institute of Theoretical Physics and Astronomy, Vilnius University}

\begin{document}

\maketitle

\begin{abstract}
We consider models of the population or opinion dynamics which result in the non-linear stochastic differential equations (SDEs) exhibiting the spurious long-range memory. In this context, the correspondence between the description of the  birth-death processes as the continuous-time Markov chains and the continuous SDEs is of high importance for the alternatives of modeling. We propose and generalize the Bessel-like birth-death process having clear representation by the SDEs. The new process helps to integrate the alternatives of description and to derive the equations for the probability density function (PDF) of the burst and inter-burst duration of the proposed continuous time birth-death processes.
\end{abstract}

\section{Introduction}
the birth-death processes or the continuous time Markov chains are of great interest in the modeling of biological and social systems \cite{Metzler2014WorldScientific}. First of all, birth-death processes are useful in demography, population dynamics, genetics, epidemic dynamics, ecology, queuing theory, \cite{Sundarapandian2009PHILearning, Lloyd2001Science,Gonzalez2008Nature}. In case of the global agent interactions or the randomly generated networks such systems exhibit the continuing stochastic fluctuations in the collective behavior \cite{Alfarano2009Dyncon,Kononovicius2016PhysLettA} with power-law in the first and the second order statistics \cite{Gabaix2009ARE,Ruseckas2011EPL,Kononovicius2012PhysA} and thus are of great importance in the finance and other social systems \cite{Alfarano2005CompEco,Gontis2006JStatMech}. 

We aim to find a reliable method of how to discriminate macroscopic behavior of such Markov systems from alternative one exhibiting the true long-range memory properties such as in the fractional Brownian motion. The main idea is to analyze PDF of the burst and inter-burst duration, which is invariant for the various non-linear transformations of the observed stochastic time series \cite{Gontis2017PhysA,Gontis2017Entropy,Gontis2018PhysA}. This opportunity is given by well defined general power-law form of PDF $P(\tau)\sim \tau^{2-H}$ for the burst and inter-burst duration $\tau$ of the stochastic time series with Hurst exponent $H$ \cite{Ding1995fbm}. For the Markov time series $H=1/2$ thus the exponent of PDF has the value $3/2$ general for all one-dimensional stochastic processes with uncorrelated increments.

The scientific uncertainty of these power-laws is related to the divergence of PDF and its moments. Fortunately, this divergence is rather formal mathematical, and one can overcome this problem dealing with a finite number of agents or a finite number of states. Indeed, an infinite number of agents or the continuous limit of the stochastic processes is a mathematical abstraction. Systems we aim to model usually have a limited number of agents and thus behave as the discrete stochastic processes. Thus we consider a continuous time birth-death process with discrete states $\lbrace0,1,2,3,...,m,...,N\rbrace$ or equivalently a system of N agents with possible two states of individual agents $\lbrace0,1\rbrace$.

The problem of the burst and inter-burst duration in such system is just a part of the general first passage time theory \cite{Gontis2017Entropy,Gontis2018PhysA}, where the moments of the first passage time for the birth-death processes have closed-form expressions \cite{Jouini2008MMOR}. Nevertheless, the explicit form of the first passage time PDF is known for a few cases and usually has the infinite sums of special functions \cite{Sasaki2009JMathPhys}. The Bessel process, written as the one-dimensional SDE has an explicit solution for the first passage time PDF. We will integrate this knowledge introducing a Bessel-like birth-death process and will show that the solution for the continuous process can be used to approximate the burst and inter-burst duration PDF of the proposed Bessel-like birth-death process.

In the second section, we discuss the relationship between the birth-death processes and the one-dimensional SDEs. Then we introduce a Bessel-like birth-death process and derive PDF of the burst and inter-burst duration. Finally, we discuss the results and make some conclusions.   

\section{the birth-death processes and the continuous SDE's}
We will use the notations of the transition rate $\lambda(m,N)$ from state $m$ to $m+1$ and the transition rate $\mu(m,N)$ from state $m$ to $m-1$, $\lambda(N,N)=0$, $\mu(0,N)=0$ and both rates are $>0$ otherwise. Let us start from the  well-defined herding model \cite{Kirman1993QJE,Alfarano2005CompEco} with rates written as
\begin{equation}
    \lambda(m,N) = (N-m) (\varepsilon_1 + m) , \quad \mu(m,N) = m (\varepsilon_2 + (N-m)). \label{eq:KirmanRates}
\end{equation}
Note, that the herding model is in the background of our approach to the modeling of stochastic volatility in the financial markets \cite{Gontis2016PhysA,Gontis2018PhysA}.
The master equation for PDF of the macroscopic state evolution $P(m,t)$ can be written using the transition rates $\lambda_m$ and $\mu_m$. In the limit of a high number of agents $N$ one can define Fokker-Planck equation for continuous variable $x=m/N$ PDF, see \cite{Gontis2017Entropy} for details, or corresponding stochastic differential equation (SDE) in Ito sense:
\begin{equation}
\rmd x=\Delta x(\lambda - \mu) \rmd t + \sqrt{\Delta x^2( \lambda +\mu)} \rmd W,
\label{eq:SDE}
\end{equation}
where $\Delta x=1/N$ is the change of $x$ during one step of the birth-death process, and $W$ is the standard Wiener noise. Transition rates in general case depend on $x$ and $N$. For the herding model, the SDE is
\begin{equation}
\rmd x = \left[ \varepsilon_1 (1-x) - \varepsilon_2 x \right] \rmd
t + \sqrt{2 x (1-x)} \rmd W ,
\label{eq:xsdedimless}
\end{equation}
For the certain forms of the transition rates, when $(\lambda(x,N) +\mu(x,N))\sim N^2$, the stochastic term in Eq. \eqref{eq:SDE} does not disappear and fluctuations in the agent system are continuing: the case of non-extensive statics \cite{Ruseckas2011PhysRevE}, having an example with the herding transition rates \cite{Ruseckas2011EPL,Kononovicius2012PhysA}. 

We consider here the burst and inter-burst duration introduced in the previous work \cite{Gontis2012ACS}. In other words, the burst duration means here the first passage time to the threshold starting from the first state above the threshold and the inter-burst duration starting from the first state below the threshold. Statistical properties of the burst and inter-burst duration are invariant regarding the non-linear transforms of the time series when one transforms the thresholds as well. 
SDE for the population ratio $y=\frac{x}{1-x}$ can be written as,
Eq. (17) from \cite{Kononovicius2012PhysA}:
\begin{equation}
\mathrm{d}y = \left[\varepsilon_{1}/y+\left(2-\varepsilon_{2}\right)\right] y \left(1+y\right) \mathrm{d} t+ \sqrt{2y}\left(1+y\right) \mathrm{d}W.\label{eq:sdey}
\end{equation}
Note that this SDE exhibits spurious long-range memory and is invariant for the transformation of variable $y\rightarrow\frac{1}{y}$  \cite{Gontis2017Entropy}. By this transformation bursts become inter-bursts and vice versa.

In case of the continuous time discrete state birth-death process, the inter-burst duration is equivalent to the first passage time from state $m-1$ to state $m$ and the burst duration starting from state $m+1$ to state $m$. The correspondence of the time (burst or inter-burst duration) between continuous and discrete descriptions is the main idea of this contribution. It will help to solve the problem of the burst and inter-burst duration statistics for the non-extensive Bessel-like birth-death process. 

After Lamperti transformation $z(y)=\sqrt{2}\arctan{\sqrt{y}}$ and assuming $\varepsilon_{1}=\varepsilon_{2}=\varepsilon$ the herding model can be written as
\begin{equation}
\rmd z = \frac{2 \varepsilon-1}{\sqrt{2}} \cot {(z\sqrt{2})}  \rmd
t + \rmd W ,\label{eq:sdezKirman}
\end{equation}
As we deal with number of agents $N$, the continuous and the discrete variables are defined in the following intervals: $0\leq x \leq 1$; $0\leq y \leq \infty$; $0\leq z \leq \pi/\sqrt{2}$; $0\leq m \leq N$. 
In the limit $z\rightarrow 0$ Eq. \eqref{eq:sdezKirman} becomes equivalent to the Bessel process 
\begin{equation}
\rmd z = \frac{2 \varepsilon-1}{2 z}  \rmd
t + \rmd W ,\label{eq:sdezBessel}
\end{equation}

\section{Bessel-like birth-death process}
\begin{figure}
\centering
\includegraphics[width=0.32\textwidth]{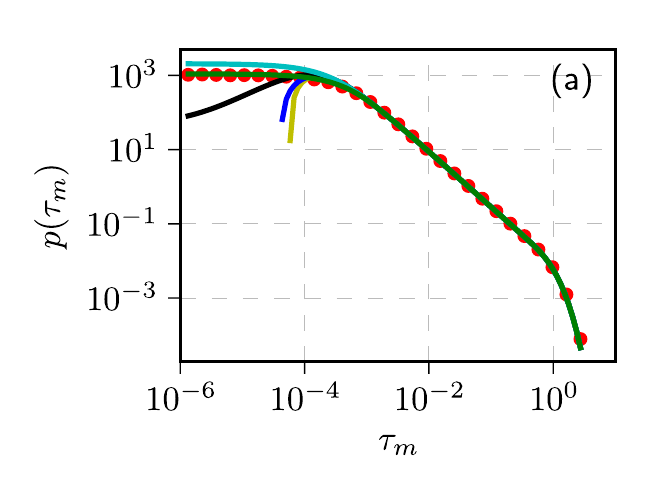}
\includegraphics[width=0.32\textwidth]{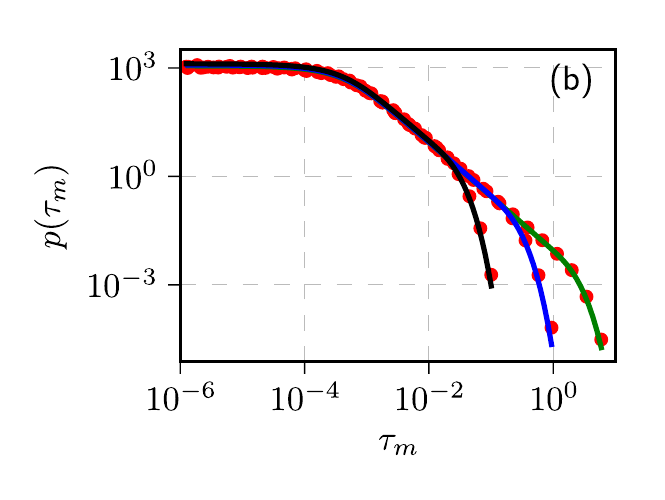}
\includegraphics[width=0.32\textwidth]{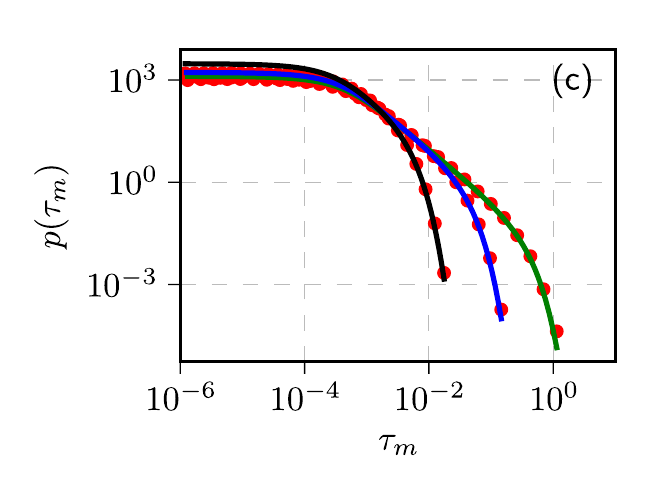}
\caption{\label{fig:Bessel}The comparison of numerical $\tau_m$ PDF (red points) with Eqs. \eqref{eq:BessTauPDF} and \eqref{eq:BessTauPDFSimp}  for the Bessel-like process \eqref{eq:BesselRates}, $N=100$. a) Eq. \eqref{eq:BessTauPDF}, $m=60$, $k_m=37$ (green line), $k_m=60$ (cyan line), $k_m=85$ (black line),$k_m=97$ (blue line), $k_m=120$ (orange line), $\varepsilon=1.5$.   b) Eq. \eqref{eq:BessTauPDFSimp}, $m=90$ (green line), $m=30$ (blue line), $m=10$ (black line), $\varepsilon=1.5$; c) Eq. \eqref{eq:BessTauPDF}, $\varepsilon=5.5$, $m=90$ (green line), $m=30$ (blue line), $m=10$ (black line).}
\end{figure}

We consider the Bessel process as an asymptotic limit of the herding model, see \cite{Kononovicius2012PhysA,Gontis2012ACS} for details. Let us to partition the interval $0\leq z \leq \pi/\sqrt{2}$ into equally spaced intervals $\Delta z=\pi/\sqrt{2}/N$ seeking to define an alternative non-extensive Bessel-like birth-death process. Transition rates of such process are as follows
\begin{equation}
    \lambda_b(m,N) = \frac{N^2}{\pi^2} \left( 1+\frac{\varepsilon-1/2}{m} \right) , \quad
    \mu_b(m,N) =  \frac{N^2}{\pi^2} \left( 1-\frac{\varepsilon-1/2}{m} \right) . \label{eq:BesselRates}
\end{equation}
Substitution of the rates \eqref{eq:BesselRates} into Eq. \eqref{eq:SDE} for $z$ gives Eq. \eqref{eq:sdezBessel}. Although the Bessel-like birth death process is not bounded, first passage times from any lower values of $m$ to the higher values of $m$ are well defined and are related to the passage time solution for the SDE \eqref{eq:sdezBessel}. Let us demonstrate the correspondence between the discrete first passage time $\tau_m$ from the state $m-1$ to the state $m$ and well defined first passage time in the continuous time SDE \eqref{eq:sdezBessel}. The PDF of passage time $\tau_m$ is given in \cite{Borodin2002Birkhauser} as
\begin{equation}
P_b^{(\nu)}(\tau_m) = \frac{z_m^{\nu-2}}{z_{m-1}^\nu} \sum_{k=1}^{k_m} \frac{j_{\nu,k} J_\nu\left(\frac{z_{m-1}}{z_m} j_{\nu,k}\right)}{J_{\nu+1}(j_{\nu,k})} \exp\left(- \frac{j_{\nu,k}^2}{2 z_m^2} \tau_m\right) ,
\label{eq:BessTauPDF}
\end{equation}
where $P_b^{(\nu)}(\tau_m)$ is PDF of the first passage times at level $z_m$ of the Bessel process with the index $\nu=\varepsilon-1$ starting from $z_{m-1}$, $J_\nu$ is the Bessel function of the first kind of the order $\nu$ and
$j_{\nu,k}$ is the $k$-th zero of $J_\nu$. The number of terms in the sum $k_m=\infty$ for the continuous Bessel process. The main discrepancy of  PDF \eqref{eq:BessTauPDF} for the continuous Bessel process and its discrete version is in the region $\tau_m \rightarrow 0$. In \figref{fig:Bessel} (a) we compare numerical calculations (Gillespie algorithm) of the first passage time PDF from the state $m=59$ to the state $m=60$ with the corresponding continuous time PDF \eqref{eq:BessTauPDF}. It is obvious that for the discrete space of the variable values there are natural limits of diffusion as the system can not move to the states lower than the state with $\mu_b(m_0,N)=0$ and for the smallest $\tau_m$ probability density approaches exponential form for direct jumps from state $m-1$ to $m$. For the Bessel-like process the space of diffusion depends on the parameter $\varepsilon$ as well. Probably there is some limit values $k_m$ of index $k$ in sum of exponential terms of Eq. \eqref{eq:BessTauPDF}, where the last term $\exp\left(- \frac{j_{\nu,k_m}^2}{2 z_m^2} \tau_m\right)$ describes the direct jumps from state $m-1$ to state $m$. We noticed that the last exponential rate $\frac{j_{\nu,k_m}^2}{2 z_m^2}$ should be equal to the biggest eigenvalue $\xi_m$ from Keilson theorem \cite{Keilson1979AMS,Gong2012JTP}. With such assumption $k_m$ can be defined from the following equation
\begin{equation}
\frac{j_{\nu,k_m}^2}{2 z_m^2}=\xi_m=2 (\lambda(m-1)+\mu(m-1)),
\label{eq:km-tauconditional}
\end{equation}
where $\xi_m$ is related to the conditional passage from state $m-1$ to $m$, not hitting the state $m-2$, see \cite{Jouini2008MMOR}. A simple equations $k_m=\frac{2m}{\pi}$ follows from Eq. \eqref{eq:km-tauconditional}, when we substitute Bessel like rates \eqref{eq:BesselRates} and use periodic property of Bessel zeros $j_{\nu,k}$.

Using the periodic property of the Bessel zeros $j_{\nu,k}=\pi k$ and the ratio $s_m=\frac{z_{m-1}}{z_m}=\frac{m-1}{m}$ we can simplify PDF \eqref{eq:BessTauPDF} as follows 
\begin{equation}
P_b^{(\nu)}(\tau_m) = \frac{s_m^{\nu+1/2}}{z_m^2} \sum_{k=1}^{k_m} \pi k \sin \left( \frac{k}{m} \pi \right) \exp\left(- \frac{(\pi k)^2}{2 z_m^2} \tau_m \right).
\label{eq:BessTauPDFSimp}
\end{equation}
For the parameter value $\nu=\varepsilon-1=0.5$ PDFs \eqref{eq:BessTauPDF} and \eqref{eq:BessTauPDFSimp} coincide. It is worth to note that the integral of normalization $S(k_m)=\int_0^{k_m}P^{(\nu)}(\tau_m) \rmd \tau_m$  is a function of $k_m$ oscillating around $1$
\begin{align}
    S(k_m) & = \frac{2}{\pi s_m^{\nu}} \sum_{k=1}^{k_m} \frac{1}{k} \sin \left( \frac{k}{m} \pi \right) = \nonumber  \\
           & \simeq \frac{2}{\pi s_m^{\nu}} \left[ \frac{(k_m-1) \pi}{m}- \frac{1}{18} \left( \frac{k_m \pi}{m} \right)^3 + \frac{1}{600} \left( \frac{k_m \pi}{m} \right)^5 \right] .
\label{eq:Normalization}
\end{align} 
Here $S(k_m)\rightarrow 1$, when $k_m\rightarrow \infty$, but the value of $k_m$ calculated from Eq. \eqref{eq:km-tauconditional} approximately is equal to the value needed for $S(k_m)$ to reach $1$ for the first time. Nevertheless, the power expansion of $S(k_m)$ in Eq. \eqref{eq:Normalization} is applicable up to $k_m=m$ and gives a slightly lower value of $k_m$ needed for the normalization. Our numerical evaluation confirms, that Eq. \eqref{eq:Normalization} is better suited to define $k_m$ for the best fit of theoretical PDF to the numerical $\tau_m$ histogram. Note that for other values of parameter $\nu\neq 0.5$ one has to integrate PDF \eqref{eq:BessTauPDF} instead of \eqref{eq:BessTauPDFSimp} in order to define normalization $S(k_m)$ and $k_m$ needed for the discrete modeling.

To confirm this form of PDF in Fig. \ref{fig:Bessel} we compare numerical calculations using Gillespie algorithm \cite{Gillespie1977JPCh} of the first passage  time $\tau_m$ (the inter-burst duration) PDF with Eqs. \eqref{eq:BessTauPDF} and \eqref{eq:BessTauPDFSimp} having only $k_m$ exponential terms in the sum. 

Finite number $k_m$ of the exponential terms in PDF \eqref{eq:BessTauPDF} explains the PDF behavior when $\tau_m\rightarrow 0$.  Differences between the continuous Bessel process and the Bessel-like birth-death process with the finite number of agents $N$ are important in this region. In both cases we have well-defined first passage time $\tau_m$ PDF, coinciding for the all $\tau_m$ values with the exception in the limit for very small values. This discrepancy appears because only direct jump $m-1 \rightarrow m$ is possible for the birth-death process when in the continuous diffusion case there is an infinite number of states in between.   

We do seek to extend the applications of Eq. \eqref{eq:BessTauPDF} for the wider class of birth-death processes and first of all for the cases with bounded diffusion. Let us add the additional terms $-\frac{\varepsilon_2-1/2}{N-m}$ to the birth rate $\lambda_b$ and $\frac{\varepsilon_2-1/2}{N-m}$ to the death rate $\mu_b$ Eq. \eqref{eq:BesselRates}. This defines the bounded Bessel-like birth-death process
\begin{align}
\lambda_{bb}(m,N) & = \frac{N^2}{\pi^2}\left(1+\frac{\varepsilon_1-1/2}{m}-\frac{\varepsilon_2-1/2}{N-m}\right) , \nonumber \\
\mu_{bb}(m,N) & =  \frac{N^2}{\pi^2}\left(1-\frac{\varepsilon_1-1/2}{m}+\frac{\varepsilon_2-1/2}{N-m}\right), \label{eq:BesselBoundedRates}
\end{align}
which is equivalent to the Bessel-like process Eq.\eqref{eq:BesselRates}, when $m \ll N$. It is useful to write the continuous SDE corresponding to this new non-extensive birth-death process, for simplicity in the case $\varepsilon=\varepsilon_1=\varepsilon_2$,
\begin{equation}
\rmd z = \frac{(\pi/\sqrt{2}-2z)(\varepsilon-1/2)}{z(\pi/\sqrt{2}-z)}  \rmd
t + \rmd W.
\label{eq:sdezBoundBessel}
\end{equation}
Compare Eq. \eqref{eq:sdezBoundBessel} with Eq. \eqref{eq:sdezKirman}, both describe the Brownian particle in symmetric potential well.

Introducing the new birth-death rates Eq. \eqref{eq:BesselBoundedRates} we generalize the Bessel-like birth-death process Eq. \eqref{eq:BesselRates}. This bounded version has a clear relation with the classical Bessel process and represents the case of diffusion in the same interval $0 \leq z \leq \pi/\sqrt{2}$ as the corresponding transformation of the herding model Eq. \eqref{eq:sdezKirman}. We do expect to adjust results of the first passage and bursting time PDF, described above, for this generalized version of birth-death process. 

First, let us make the numerical comparison of the inter-burst duration $\tau_m$ PDF calculated with the bounded and unbounded versions of the Bessel-like birth-death process.
\begin{figure}
\centering
\includegraphics[width=0.45\textwidth]{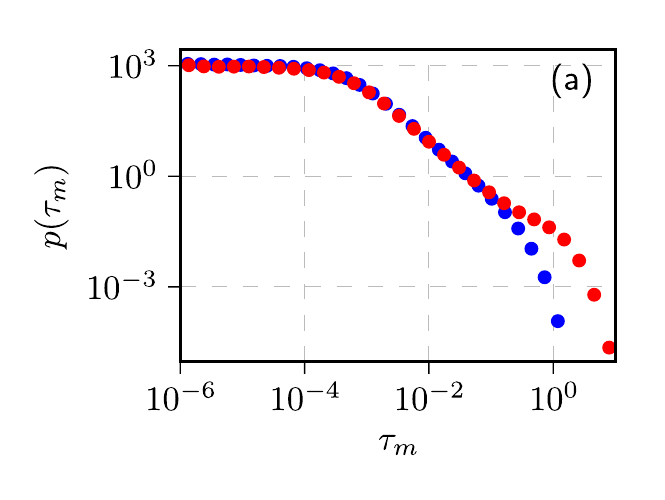}
\includegraphics[width=0.45\textwidth]{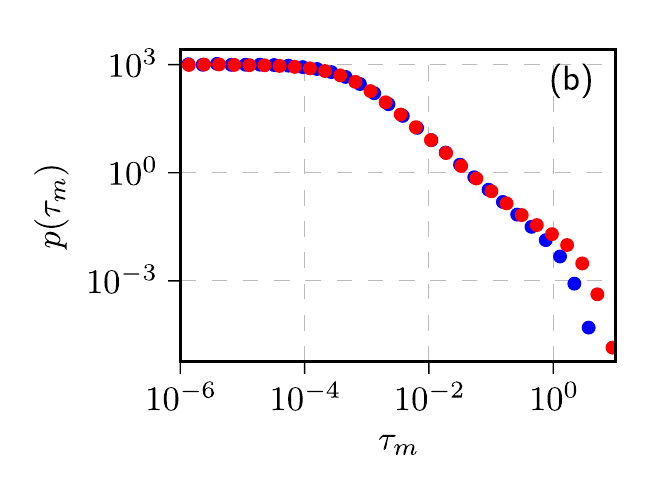}
\caption{\label{fig:BoundedBessel}The comparison of numerical inter-burst time $\tau_m$ PDF for the Bessel-like birth-death process (blue points) with the bounded Bessel-like process (red points). Parameters are as follows: $N=100$, $m=70$, a) $\varepsilon=3.5$, b)  $\varepsilon=1.5$.}
\label{fig:2}
\end{figure}
The numerical comparison of the bounded and unbounded versions, see examples in \figref{fig:2}, reveals that differences of PDF appear only for the tail part representing the exponential cut off of the distribution. In order to account for this peculiar  behavior for the bounded version of the Bessel-like birth-death process we propose to add one more exponential term to the PDF of $\tau_m$ given by Eq. \eqref{eq:BessTauPDF} with the number of exponential terms $k_m$ estimated from the normalization. 
\begin{equation}
P_{bb}(\tau_m)= (1-\rho) P_b(\tau_m) + \frac{\rho}{
\tau_{m0}} \exp\left(-\frac{\tau_m}{\tau_{m0}}\right).
\label{eq:boundedPDF}
\end{equation}
Here are two parameters in this new form of PDF. $\rho$ defines the weight of the new exponential term and $\tau_{m0}$ is a scale of exponential cut off. We define these parameters from the first $\tau_{m,1}$ and second $\tau_{m,2}$ moments of the first passage time, calculated from the rates of birth-death process, see \cite{Jouini2008MMOR}. Two equations for the needed parameters are as follows
\begin{align}
    \tau_{m,1} & = (1-\rho) Q_1 + \rho \tau_{m0}, \nonumber\\ 
    \tau_{m,2} & = (1-\rho) Q_2 + 2 \rho \tau_{m0}^2. \label{eq:BoundedPDFparameters}
\end{align}
Here $Q_1$ and $Q_2$ are the coefficients defined as sums of the Bessel functions.
\begin{align}
    Q_1 &= 4 z_m^2 \left( \frac{m}{m-1} \right)^{\nu} \sum_{k=1}^{k_m} \frac{J_\nu\left(\frac{m-1}{m} j_{\nu,k}\right)}{j_{\nu,k}^3 J_{\nu+1}(j_{\nu,k})}, \nonumber\\
    Q_2 &= 16 z_m^4 \left( \frac{m}{m-1} \right)^{\nu} \sum_{k=1}^{k_m} \frac{J_\nu\left(\frac{m-1}{m} j_{\nu,k}\right)}{j_{\nu,k}^5 J_{\nu+1}(j_{\nu,k})}.
\label{eq:BoundedPDF-Q12}
\end{align}

\begin{figure}
\centering
\includegraphics[width=0.45\textwidth]{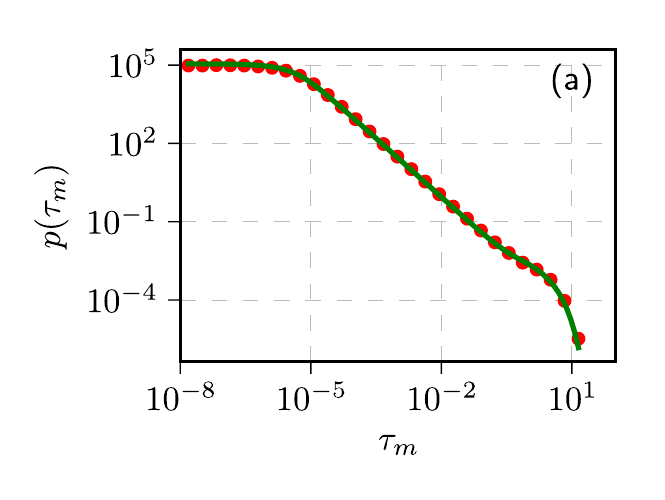}
\includegraphics[width=0.45\textwidth]{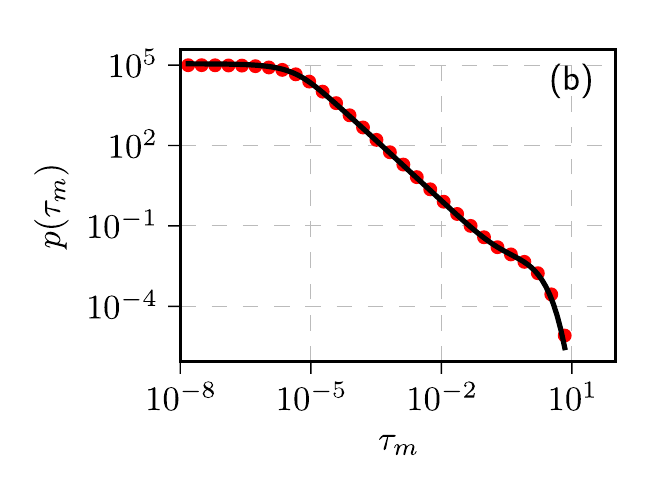}
\caption{The comparison of numerical inter-burst duration $\tau_m$ PDF (red points) with Eq. \eqref{eq:boundedPDF} for the bounded Bessel-like process \eqref{eq:BesselBoundedRates}, $N=1000$,  a) $\varepsilon=1.5$, $m=800$ (green line);   b)  $\varepsilon=3.5$, $m=700$ (blue line).}
\label{fig:3}
\end{figure}

We confirm this approximation by the comparison in Fig. \ref{fig:3} of the numerical inter-burst duration PDF (Gillespie algorithm) for the bounded Bessel-like birth-death process Eq. \eqref{eq:BesselBoundedRates} and proposed PDF \eqref{eq:boundedPDF} with $\rho$ and $\tau_{m0}$ evaluated from Eqs. \eqref{eq:BoundedPDFparameters}.

\section{Discussion and conclusions}
The relation between the stochastic and agent-based description of social systems is of great interest. The most direct connection is possible through the correspondence of the birth-death processes to the one-dimensional SDEs. Being a well defined mathematical tool in population and opinion dynamics birth-death processes can be easily treated as an outcome of the heterogeneous agent system with two opinions. The macroscopic description of such social systems results in nonlinear SDEs exhibiting the first and the second order power-law statistics \cite{Alfarano2005CompEco,Gontis2012ACS,Ruseckas2012ACS}. The herding model serves as the most popular birth-death process used in the contemporary models of social systems \cite{Alfarano2005CompEco}.
Here we proposed the Bessel-like birth-death process as an alternative to the herding model and demonstrated that both are asymptotically related. the Bessel-like birth-death process is a convenient, discrete version of modeling having the well established corresponding continuous version of the Bessel process Eq. \eqref{eq:sdezBessel}.      We explored this correspondence, first of all, for a better understanding of the statistical properties of the burst and inter-burst duration. If in the continuous description PDFs of the burst and inter-burst duration are divergent for the small time intervals, PDFs  are normalized and well defined for the birth-death processes. We do expect that the proposed simple form of PDF \eqref{eq:BoundedPDF-Q12} can be very useful for the empirical analyses of the time series exhibiting power-law statistical properties and helpful defining whether a process has the spurious or the true long-range memory. The preliminary empirical study of the absolute return and the trading activity time series in the financial markets \cite{Gontis2017PhysA} has confirmed the power-law exponent $3/2$ for the burst and inter-burst duration PDF. Furthermore, the numerical experiments \cite{Gontis2018PhysA} with earlier proposed the Consentaneous agent based and stochastic model of the financial markets \cite{Gontis2014PlosOne} revealed that the Markov chain of the agent herding interactions may be in the origin of this power-law. Here we extend the scope of the birth-death processes capable to exhibit the  power-law statistical properties in the macroscopic description.

Indeed, the continuous SDE \eqref{eq:sdezBoundBessel} after variable transform $y=\frac{z}{\pi/\sqrt{2}-z}$ and time scaling $t_s=\frac{2 t}{\pi^2}$, see for comparison the variable transformations used with the herding model Eq. \eqref{eq:sdey}, can be written as
\begin{equation}
\mathrm{d}y = \left[\frac{\varepsilon-\frac{1}{2}}{y}- \left( \varepsilon-\frac{3}{2} \right) \right]\left(1+y\right)^3 \mathrm{d}
t_s+ \left(1+y\right)^2 \mathrm{d}W_s.
\label{eq:sdeyBessel}
\end{equation} 
This proves that the Bessel-like birth-death process with a sufficiently high number of agents $N$ may generate time series having the same power-law statistics as defined by the class of non-linear SDEs, considered in the series of papers \cite{Kaulakys2005PhysRevE,Ruseckas2011PhysRevE,Ruseckas2012ACS,Ruseckas2014JStatMech}
\begin{equation}
\mathrm{d}x=\left( \eta-\frac{\lambda}{2} \right) x^{2\eta-1}\mathrm{d} t_s +x^{\eta} \mathrm{d} W_s
\end{equation} 
The SDEs belonging to this class generate stationary power-law PDF of the variable $P(x)\sim x^{-\lambda}$ with the exponent $\lambda$ and power spectral density $S(f)\sim \frac{1}{f^\beta}$ with the exponent $\beta=1+\frac{\lambda-3}{2\eta-2}$. For the Bessel-like process Eq. \eqref{eq:sdeyBessel}, $\eta=2$ and $\lambda=2 \varepsilon+1$.

the observed power-law statistical properties can be easily confused with the true long-range memory.   the Bessel-like birth-death process presented here, previously considered the herding model and the other birth-death processes \cite{Kononovicius2019JSM} probably can be used for the modeling of the social systems exhibiting continuing fluctuations and non-extensive statistics  \cite{Gontis2009AIP,Kononovicius2014EPJB}. All these Markov chains may generate very similar statistical properties of the time series and should have very similar PDFs of the burst and inter-burst duration. the presented correspondence between the birth-death processes and the class of non-linear SDEs can be very helpful in building the theoretical background of various social systems exhibiting power-law behavior. 

The empirical analyses of the burst and inter-burst duration statistics for various social systems should be very helpful searching for the best alternatives for the modeling. The huge amount of the data available for the financial markets makes this social system as the most promising in such research projects. From our point of view, it is worth to analyze the time series of demand and supply summarized in the limit order books. Nevertheless, we have to recognize that such research will be effort and time consuming as one has to deal with the limit order data in a period of a few years. 



\end{document}